\documentclass[preprint,12pt]{elsarticle}

\usepackage{amssymb}
\usepackage{amssymb}
\usepackage{amstext}
\usepackage{gensymb}
\usepackage[version=4]{mhchem}
\usepackage[super]{nth}
\usepackage{comment}
\newcommand{\um}{$\mu$m }

\newcommand{\atp}{at.\%}

\usepackage{lineno}

\journal{Acta Electrochemica}

\begin{document}

\begin{frontmatter}



\title{Development of  an automated millifluidic platform and data-analysis pipeline for rapid electrochemical corrosion measurements: a pH study on Zn-Ni}

\author[mmsd]{Howie Joress (0000-0002-6552-2972) \corref{cor}\corref{equal}}
\cortext[equal]{These authors contributed equally to this work}
\cortext[cor]{Corresponding author}

\ead{howie.joress@nist.gov}

\address[mmsd]{Materials Measurement Science Division, National Institute of Standards and Technology, Gaithersburg, MD 20889, USA}
\author[mmsd]{Brian DeCost(0000-0002-3459-5888)\corref{equal}}

\author[UW,mmsd]{Najlaa Hassan (0000-0003-3506-0057)}
\address[UW]{Department of Engineering Physics,University of Wisconsin-Madison, Madison, WI 53708}
\author[MSED]{Trevor M. Braun (0000-0002-9779-3785)}
\address[MSED]{Materials Science and Engineering Division, National Institute of 
 Standards and Technology, Gaithersburg, MD 20899}
\author[mmsd]{Justin M. Gorham (0000-0002-0569-297X)}
\author[mmsd,UofT]{Jason Hattrick-Simpers (0000-0003-2937-3188)}
\address[UofT]{Department of Materials Science and Engineering, University of Toronto, Toronto, ON M5s 1a4, Canada}
\begin{abstract}
We describe the development of a millifluidic based scanning droplet cell platform for rapid and automated corrosion.  This system allows for measurement of corrosion properties (e.g., open circuit potential, corrosion current through Tafel and linear polarization resistance measurements, and cyclic voltammograms) on a localized section of a planar sample.  Our system is highly automated and flexible, allowing for scripted changing and mixing of solutions and point-to-point motion on the sample.  We have also created an automated data analysis pipeline.  Here we demonstrate this tool by corroding a plate of electroplated \ce{Zn85Ni15} alloy over a range of pH values and correlate our results with XPS measurements and literature. 
\end{abstract}




\end{frontmatter}




\section{Introduction}
Recently there has been great interest in application of machine learning to materials science research\cite{stein2019progress}.  These types of approaches are particularly important in areas such as corrosion science, where the underlying physics is complex and difficult to model using traditional methods\citep{scully2019future}.  One difficulty in applying these methods to corrosion is the lack large data sets that can be used to train such ML models.  To this end, we have worked to develop an automated platform for rapidly acquiring electrochemical properties of materials, specifically corrosion, along with an automated data analysis pipeline.  Specifically, we are able to measure electrochemical corrosion properties on planar samples with a variation in composition or structural properties across the surface.

\citet{muster2011review} provides an excellent review of high-throughput electrochemistry methods and categorizes various approaches.  Our system is an example of a scanning probe system, specifically a scanning droplet cell (SDC), that can localize the electrolyte to a single spot on the sample, with measurements being done in series.  A review of early SDCs is provided by \citet{lohrengel2000electrochemical}. More modern implementations include a mechanism for ensuring flow through the droplet to minimize saturation or depletion of ions near the surface\cite{klemm2011high,gregoire2013scanning}.  Advantages of an SDC include minimal sample geometrical constraints or preparation (for instance, there is no need to pattern electrical contacts onto the sample), minimal interaction of the electrolyte with the surface away from probed region, and low electrolyte consumption.

In this work we describe our fully automated SDC based platform that includes a  millifluidic system to supply electrolyte to the cell head.  Because many materials of interest to us are hydrophilic, an o-ring is used to localize the electrolyte to a specific portion of the sample in contact with the cell head.  Electrolyte flows through the cell head which contains reference and counter electrodes, and the sample acts as the working/sensing electrode, creating a complete 3 electrode electrochemical cell.  The system automation includes sample manipulation, fluid handling, and data collection and reduction.  Our system includes a recirculating loop for the electrolyte to allow for relatively high flow rates while minimizing fluid waste.  Further, our set-up allows for automated mixing and changing of electrolyte solution between points. Here we discuss the details of the system as well as demonstrate its use in probing the corrosion properties of a Zn-Ni alloy. Zn alloys are of great technological interest as they are frequently used in corrosion prevention coatings in a variety of industries including the automotive industry.  Alloying the Zn with other metals, particularly Ni in the 11 \atp{} - 15 \atp{} range, is known to greatly decrease the corrosion rate of the material by as much as a factor of 5 times in a neutral salt spray test when compared to pure Zn through a compositionally as well as a structurally mediated mechanism\citep{park2020improvement, gavrila2000corrosion,ghaziof2014electrodeposition,verberne2003zinc}.

\section{System design}
\begin{figure*}
    \hspace*{-2.0cm}
    \centering
    \includegraphics[width=170 mm]{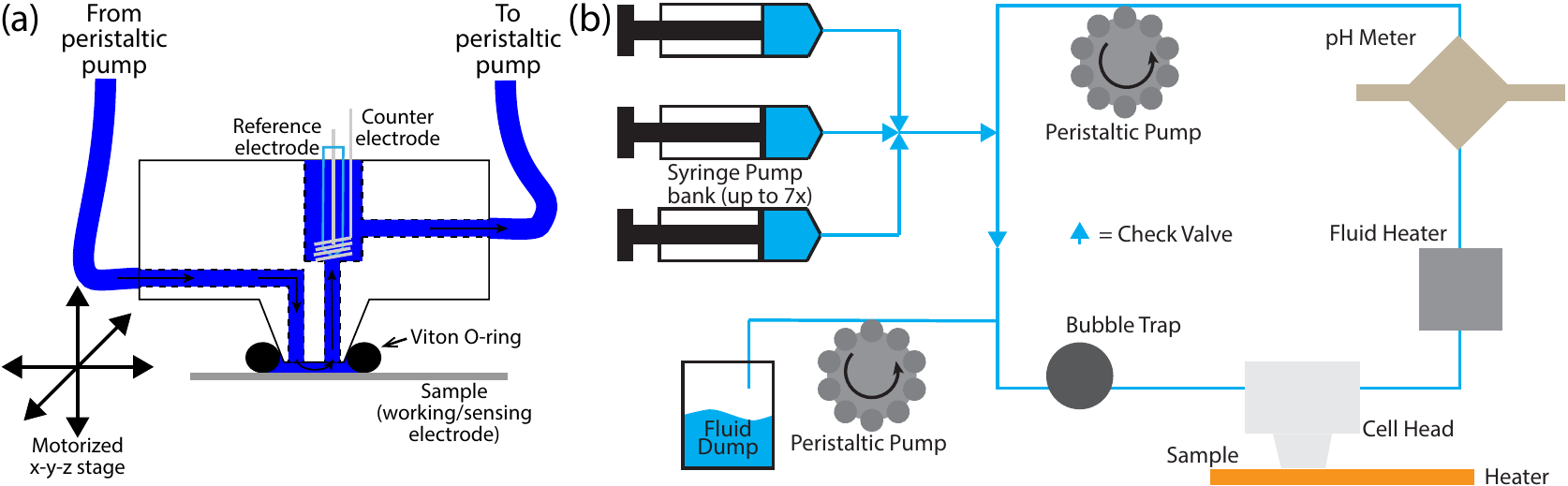}
    \caption{Representative schematics of the SDC system (not to scale).  (a) is a closeup of the cell-head showing the flow of the fluid through the cell.  The fluid enters the cell on the left, passing through the cell block to the o-ring defined volume between the cell and the sample.  The electrolyte then travels back up through the cell passing by the reference electrode and the counter electrode.  (b) shows the fluid handling system including the syringe pumps  (simplified here to three sources).  Arrowheads are physical check-valves in the system.}
    \label{fig:sdc_schem}
\end{figure*}

The SDC platform comprises 4 main components: 1) the cell head, 2) sample holder and manipulator, 3) fluid handling system, 4) control computer and electronics.  The cell head, illustrated schematically in Fig. \ref{fig:sdc_schem}(a), is machined out of a PTFE (polytetrafluoroethylene) block (a machine drawing is included in the SI).  Fluid flows into the cell through an internal channel then out into a volume between the cell head and the sample, defined by a flouroelastomer o-ring (SAE -005).  This volume, with a $\approx 4.5$ mm diameter contact area on the sample, is where the electrochemistry occurs; the sample acts as the working/sensing electrode (WE).  The electrolyte then flows back up into the cell head where it enters a chamber containing a standard commercial Ag/AgCl reference electrode (RE, with saturated KCl electrolyte) and several centimeters of 0.5 mm Pt wire, acting as the counter electrode (CE), wrapped around it.  The surface area of the wire is at least an order of magnitude greater than that of the active area of the working electrode.  After passing by these electrodes the fluid leaves the cell head to return to the fluid handling system.  

Together the RE, CE, and WE as described above constitute a conventional three-electrode electrochemical cell.  We note that this cell design has a relatively high resistance caused by the distance and small flow channel between the working electrode and the reference and counter electrodes.  We have endeavored to minimize this source of resistance to the extent possible given geometrical and machining limitations.  For many measurement types and samples, the current will be low enough that the voltage drop between the sets of electrodes will be minimal and can be accounted for in post-processing.

The fluid handling system, shown schematically in Fig \ref{fig:sdc_schem}(b), is built around a recirculating loop constructed of ultra-chemical resistant Tygon tubing\footnote{\textbf{Certain trade names and company products are mentioned in the text or identified. In no case does such identification imply recommendation or endorsement by the National Institute of Standards and Technology (NIST), nor does it imply that the products are necessarily the best available for the purpose.}} with a 1/16 in (1.6 mm) inner diameter (ID).  The loop is driven by a peristaltic pump (Ismatec Reglo ICC). Prior to entering the cell head the fluid passes through a bubble trap to prevent entrapped air from entering the cell head and affecting the measurement.  Downstream of the cell head, the fluid passes through a custom inline pH sensor (Microelectrodes inc.).  The cell head and the rest of the system components were constructed with an aim of minimizing fluid volume, with the volume being ~4 mL.

The system is filled from a bank of syringe pumps acting as source reservoirs for a set of stock solutions.  The flows from each syringe pumps is merged and directed into the main fluid loop, with the fluid flow direction ensured by a set of check valves. The syringes can be simultaneously pumped at different pump rates to alter the composition of the electrolyte in the loop.  Used electrolyte is pumped out of the loop via a second peristaltic pump channel into a fluid.  Refilling of the loop is typically done by pushing fluid from the syringes at approximately  (with a small positive bias) the same rate as fluid is pulled from the loop into the dump.  The peristaltic pump in the loop is reversed such that fluid must first pass through the cell head and the majority of the loop before being dumped.  At the end of each fill cycle the pump driving the loop is stopped so that the remainder of the loop can be purged of used electrolyte.

The cell head is mounted on a vertical stage to lower it so such that it is in contact with the sample and to raise it to allow the cell head to move laterally relative to the sample. The head is attached to this stage via a vertically spring-loaded plate to prevent excess stress on the sample and to accommodate misalignment between the sample surface and the cell head o-ring.

The sample is mounted on a 3 point kinematic mount.  The sample holder itself consists of a silicone heating pad pressed between two metal plates along with insulation underneath to direct the heat upward towards the sample.  Between the upper plate and the sample is a thin PTFE sheet to protect the sample holder from excess electrolyte and to electronically isolate the sample.  The sample is held to the plate with plastic screws which push copper clips onto the sample surface for electrical and mechanical contact.  There is a stack of two perpendicular, horizontal, linear motion stages under the kinematic mount to precisely locate specific portions of the sample under the cell head.

The SDC system is designed to collect data in an automated fashion. A control program executes experimental protocols that specify sample coordinates, associated electrolyte chemistry and a series of electrochemical measurements and their parameters. Experimental protocols can be executed from a pre-planned script or dynamically generated at runtime based on the results of online data analysis. Each measurement sequence proceeds as follows:  (1) The cell head is lifted off the sample.  (2) The sample is then translated to place the cell head over the correct position on the sample.  (3) An external needle then sprays deionized (DI) water on the local area to clean the surface.  (4) Fresh electrolyte is then pushed into the cell head creating a droplet, wetting the sample surface.  (5) The cell head is then brought down into contact with the sample, with excess fluid being pulled through the cell head and an external suction needle. (6) Once the cell head is in place, fresh electrolyte solution is purged through the system turning over the solution volume in the loop 3-5 times.  (7) The control program then executes the sequence of electrochemistry measurements and performs streaming diagnostic analyses for data quality monitoring and adaptive control capability.  (8) Following the measurements the cell head lifts up slightly to allow solution to be pulled out of the cell head before it is moved.

\section{Experimental}
For this study we characterized a steel plate electroplated with a Zn-Ni alloy.  Using X-ray fluorescence we determined that the coating had a stoichiometry of \ce{Zn85Ni15} and was about 85 \um thick.  We also characterized the phase of the alloy using x-ray diffraction, finding the film was predominately in the $\gamma$ phase, \ce{Zn11Ni2}, with a trace amount of the $\delta$ phase, \ce{Zn22Ni3}, as expected from the phase diagram\cite{nash1987ni}.

We performed the electrochemical corrosion measurements using a mixture of electrolytes to vary the pH from acidic to basic.  On the acidic side we combined 0.5 M \ce{H2SO4} with 0.5 M \ce{K2SO4} in varying ratios.  Similarly, for basic electrolytes, we combined 1 M KOH with 0.5 M \ce{K2SO4}.  These concentrations were chosen to keep the ion concentration roughly constant.

\begin{figure}[!htp]
    \vspace{-6em}
    \centering
    \includegraphics[width=84mm]{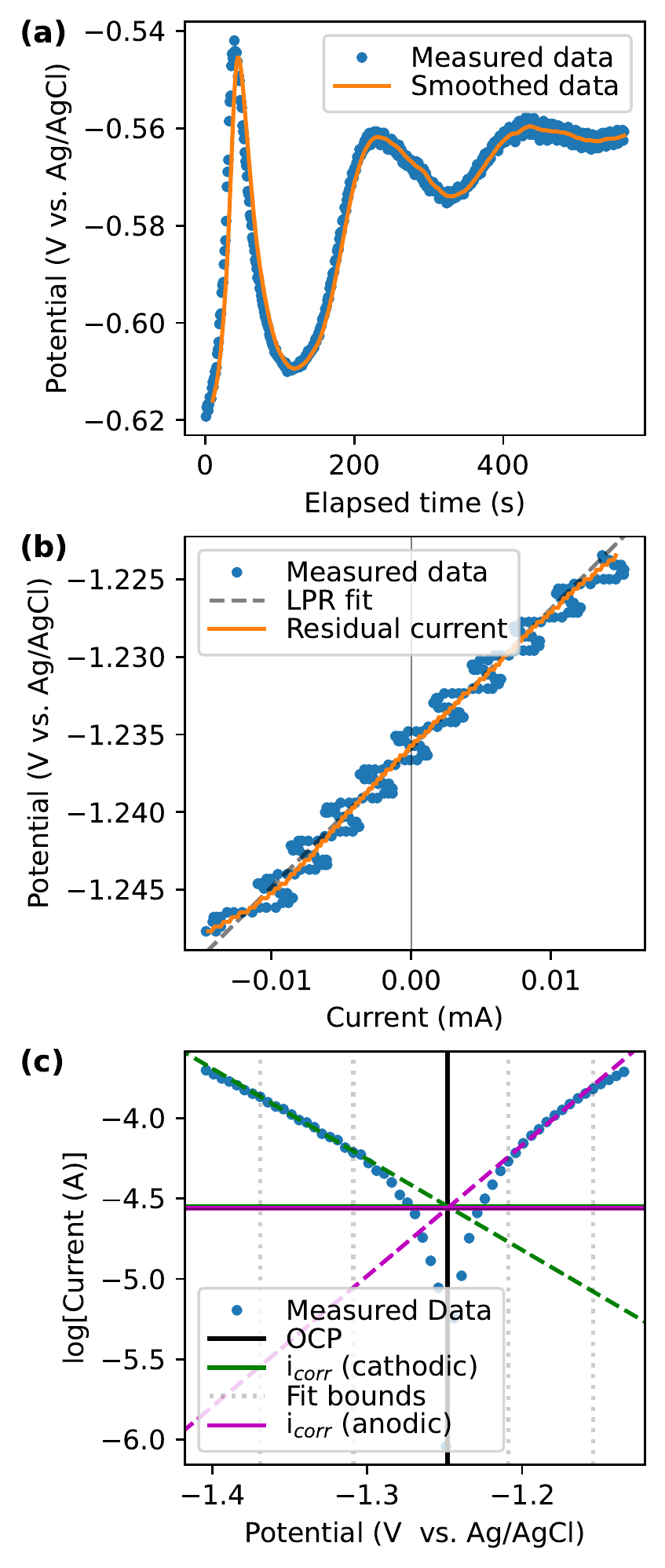}
    \caption{Representative data for each of the three measurements performed: (a) OCP hold showing the potential as a function of time.  The raw data is shown with a time averaged value and the scan stopped after the OCP was stable.  (b) shows an LPR measurement.  In this case a sine background was removed as shown.  (c) shows a Tafel measurement and the accompanying data reduction including the two fits to the Tafel slopes and the resulting corrosion current. }
    \label{fig:analysis}
\end{figure}

Every set of measurements is performed on a fresh portion of the plate surface.  At each point we performed a set of four types of electrochemical corrosion measurements, ordered from least to most destructive.  After the cell was purged we began monitoring the open circuit potential (OCP), recording it at 1 Hz.  We waited to proceed until the OCP was stable, specifically until a 10 s running average was stable to within +/- 2 mV for 90 s.  We began checking for stabilization 300 s after beginning to record the OCP and did not allow stabilization to occur for more than 1800 s.  We report the average of the OCP over the final 20 s as the OCP of the stable surface.  An example trace is plotted in Fig. \ref{fig:analysis} (a).

We then performed a series of three linear polarization resistance measurements.  These measurements swept the voltage at a rate of 0.125 mV/s over a range of +/- 12 mV, relative to the measured OCP prior to the scan, while recording the reaction current, $I$.  Since this is a relatively small deviation in potential from the OCP, these measurements are expected to have minimal effect on the surface and therefore all three duplicate measurements should be nominally identical.  Following each measurement the polarization resistance (PR) was extracted, as illustrated in Fig \ref{fig:analysis}(b) using an automated routine, as follows: The corrosion potential is identified as the data-point closest to zero current.  A line is then fit to data within +/- 5 mV of that datum to determine the PR.  In some cases an oscillation in the current was present that we attribute to the action of the peristaltic pump in the loop.  If the initial goodness of fit was insufficient ($R^2<0.95$), we first interpolate the data with a sine wave combined with a 5th order polynomial.  We then remove the oscillatory component and fit the previously described linear PR model to the polynomial component of the interpolation model.  Only PR from linear fits with an $R^2>0.95$ were kept.  Finally, we take the average of the polarization resistances (as well as the corrosion potential) over the three iterations.

Third, we perform a Tafel measurement.  These measurements are similar to the LPR scans above but we scan +/- 250 mV from the corrosion potential, at a rate of 5 mV/s.  Tafel fits can provide a measure of the corrosion current, $i_{corr}$, through the application of the Butler-Volmer model \cite{jones1996principles}. In practice extraction of $i_{corr}$ often requires expert selection of the linear region of the Tafel branches, a methodology that is incompatible with our fully automated approach.  Here we rely on a modified version of the method developed and validated by \citet{agbo2019algorithm}.  This method essentially performs a sensitivity analysis for a series of linear fits to a portion of both $I$ and the $log(I)$ as a function of applied potential; the anodic branch and the cathodic branch are analyzed separately. The fitting series systematically varies both the position and width of the fitting window. Windows with fits to either the log or linear current below a critical $R^2$ goodness-of-fit are discarded. The remaining fits were placed in bins based on their slope in log space. The bin with the greatest range of window widths is selected as the Tafel slope; we select the fit with the highest $R^2$ value within that bucket to be the Tafel fit for that branch.  A corrosion potential is also determined as the minimum voltage corresponding to the minimum in current.  An $i_{corr}$ for each branch is determined as the current of the Tafel fit at the corrosion potential.  As both branches are analyzed independently, we use the difference between the anodic and cathodic determined $i_{corr}$s as a test of goodness-of-fit, allowing no more than 0.25 decades between them.  The $i_{corr}$ for the measurement is taken as the average of those derived from each branch as demonstrated in Fig.\ref{fig:analysis}(c).

Finally, we measure a cyclic voltammogram (CV).  In this case the voltage is driven from -1.3 V to 0.5 V, cycling 2 times at a rate of 75 mV/s.  Unfortunately, this was a case where, due to the high dissolution rate, the current traveling between the surface and the counter/reference electrode became too high.  This suppressed the effective voltage at the surface of the sample, causing the CV scans to be uninterpretable, as seen in examples plotted in the SI.  This happened for all cases other than the highest pH measurements.  To that end, we do not make use of these CV measurements in this work.  Examples of these measurements can be found in our earlier work using an earlier iteration of this SDC system \cite{joress2020high}.

X-ray Photoelectron Spectroscopy (XPS) data was collected on an Axis Ultra DLD from Kratos Analytical (Chestnut Ridge, NY) using monochromatic Al K$\alpha$ X-ray source operating at 150 W.  The base pressure of the sample chamber below $10^{-7}$ Pa.  The surface was neutralized with low energy electrons and disconnected from electrical ground to minimize differential charging.  Emitted electrons were collected along the surface normal and analyzed at a 40 eV pass energy.  XPS spectra were collected from a single spot for each experimental condition from a ‘large’ area defined by a hybrid lens and slot aperture.  Spectra were acquired for the Zn 2p3, Ni 2p3, Cu 2p3, O 1s, C 1s, and S 2p regions and fit with a Shirley background (U2 Tougaard for O 1s and C1s) using CasaXPS (Teignmouth, UK).  Qualitative estimates for elemental composition were calculated from peak areas corrected using elemental relative sensitivity factors provided by the manufacturer.  Charge correction was made by shifting the Zn 2p3 feature to 1022 eV.  A portion of the data is shown in supplemental information (SI).
\section{Results and Discussion}
\begin{figure}
    \centering
    \vspace{-5em}
    \includegraphics[width=89mm]{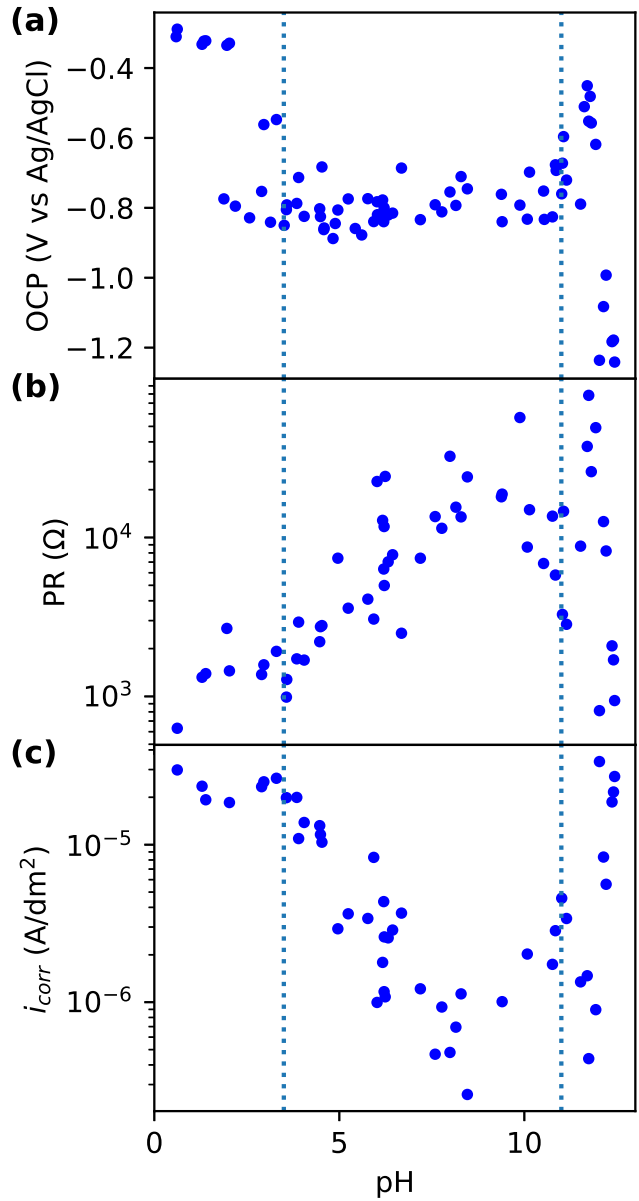}
    \caption{Results from each measurement type plotted as a function of pH.  Each point represents a scan within a measurement routine on a fresh portion of the sample.  (a) shows the final open circuit potential.  (b) shows the average polarization potential over three sweeps.  (c) shows the $i_{corr}$ extracted from the Tafel measurements.  Vertical lines are included as a guide to the eye at 3.5 and 10.5.}
    \label{fig:corr_comp}
\end{figure}
Fig. \ref{fig:corr_comp} shows the results of the various electrochemical scans plotted as a function of pH over many measurements, each point representing a single point on the sample with the measurements performed as described above.  Fig. \ref{fig:corr_comp}(a) shows the open circuit potential 
at the end of the OCP hold as a function of pH.  
We can see that at low pH (less than $\approx$ 3.5) the OCP is quite high, near -0.3 V vs. Ag/AgCl. 
As we increase the pH towards neutral there is an abrupt shift downward to approximately -0.8 V.
This OCP is maintained with a slight drift upward until a pH of about 10 is reached, where again the OCP increases until reaching a peak around 0.5 V 
before abruptly dropping down to -1.2 V at pH 11 .  

\begin{figure}
    \centering
    \includegraphics[width=90 mm]{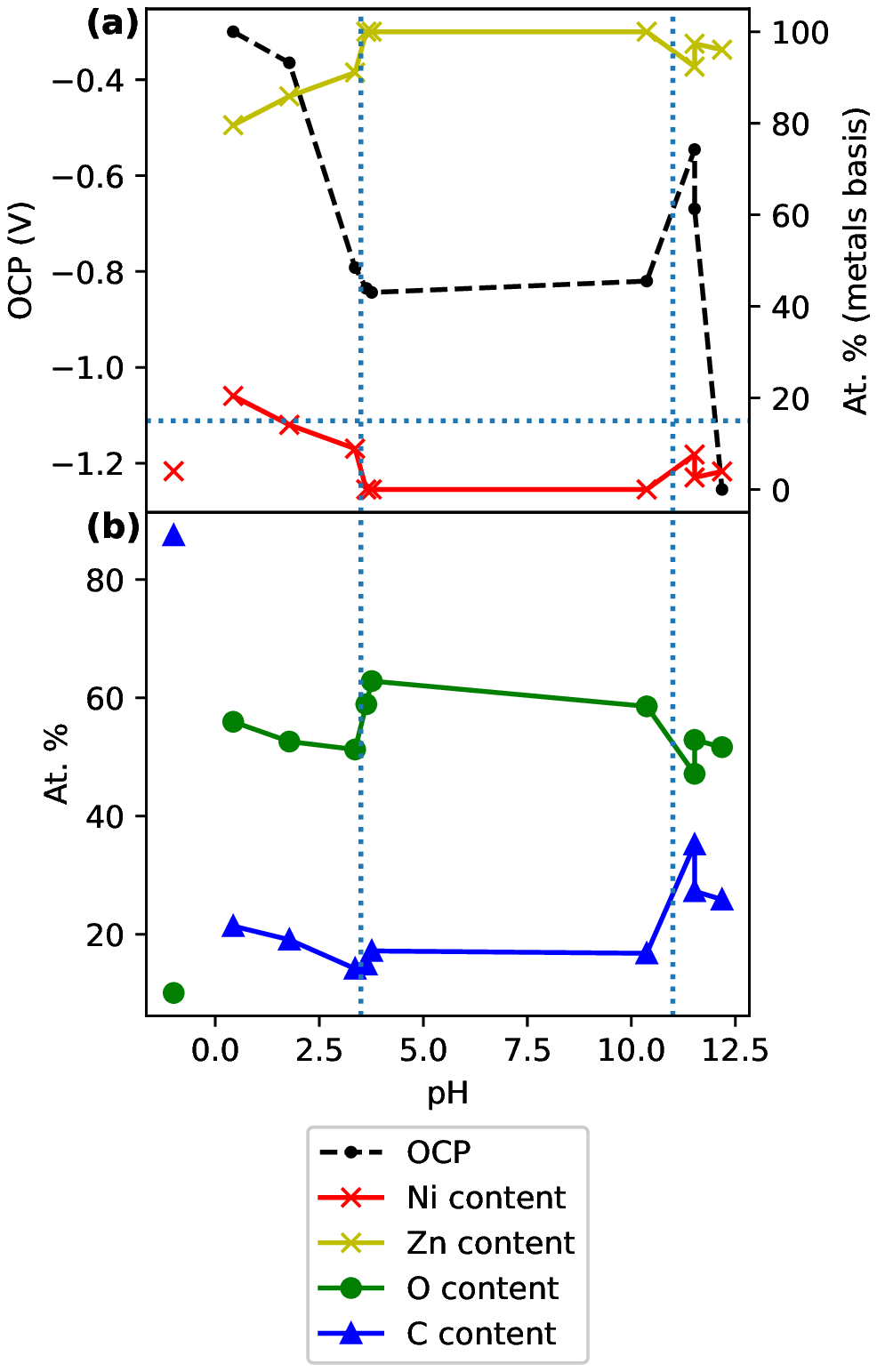}
    \caption{Compositional and chemical analysis for 9 points by ex situ XPS.  Each point represents a measurement on a unique part of the sample following OCP holds in an electrolyte whose pH varies by point. An additional point was taken on the as-received metal surface and is shown on the left as unconnected points. Vertical lines are included as a guide to the eye at pH of 3.5 and 10.5.  The Horizontal line represents the bulk Ni content for XRF.  (a) shows variation of OCP and metal fraction of Zn and Ni and (b) shows change in O and C content in absolute atomic percent}
    \label{fig:xps}
\end{figure}

Fig. \ref{fig:corr_comp} (b) and (c) show two metrics for corrosion rate: (b) shows the PR and (c) shows the corrosion current. $PR\propto 1/i_{corr}$ and both measurements correlate well with $R^2=0.78$ (See supplemental figure).  This data suggests that, starting with our most acidic measurements, the corrosion rate is high and relatively flat through  a pH of  $\approx$3.5, then begins to decrease until it reaches a minimum at a pH of  $\approx$9.5.  Moving to more basic electrolytes the corrosion rate precipitously decreases through our highest pH measurements.

As a complementary metric, we used XPS (x-ray photoelectron spectroscopy) to better understand the stable surface at OCP as a function of pH.  XPS measurements were taken at nine points, each following OCP holds, each carried out at a different pH value.  A tenth point was measured on the as-received surface for comparison.  The results are shown in Fig. \ref{fig:xps}.    The samples exposed to the lowest pH acid had the highest Ni content at 20 at.\% on a metals basis ($at.\%{} Ni/(at.\%{} Ni +at.\%{} Zn)$).  The Ni content then dropped until it was negligible at around a pH of 3.5.  On the basic side, the point closest to neutral still had negligible detectable Ni.  The content then rose to 8 \atp{} metals basis before falling back down for the last two points to 3 \atp{} and 4 \atp{} respectively.  Oxygen accounts for 48 \atp{} - 62 \atp{} for all points.  C content varied across the pH series with it starting at 21 \atp{} at the low pH end, before dipping down to 15 \atp{} in the middle of the range, then rising to 35 \atp  where the Ni content increased before falling again.    XPS can also provide some chemical information: Zn was particularly hard to identify from photoelectron spectra alone (see SI Figure).  Modified Auger parameters were extracted from survey spectra to be roughly at 2009.6 eV (data not shown), which could be consistent with  \ce{ZnCO3} or ZnO \cite{SRD20,Winiarski2018,dake1989auger}.  Similarly the Ni was predominantly in a hydroxide-like electronic configuration for all but the highest pH point, which was more consistent with zero valent Ni.  When present, Ni 2p could be identified as \ce{Ni(OH)2} due to the characteristic photoelectron line (at ~856.1 eV) coupled with a satellite shakeup broadly centered between 861-862 eV  \cite{dube1995electrodeposition,SRD20} while missing the multiplet splitting commonly observed with NiO.
The control spot had 4 at.\% Ni on a metals basis with 10 \atp{} of O. The C was strongest at 88 \atp{} for the unexposed spot, though presumably some of this C is from atmospheric hydrocarbons adhering to the surface.  In this case the Ni signal was consistent with metallic and the Zn configuration was indeterminate due to the strong C content.

\begin{figure*}
    \centering
    \includegraphics[width=140 mm]{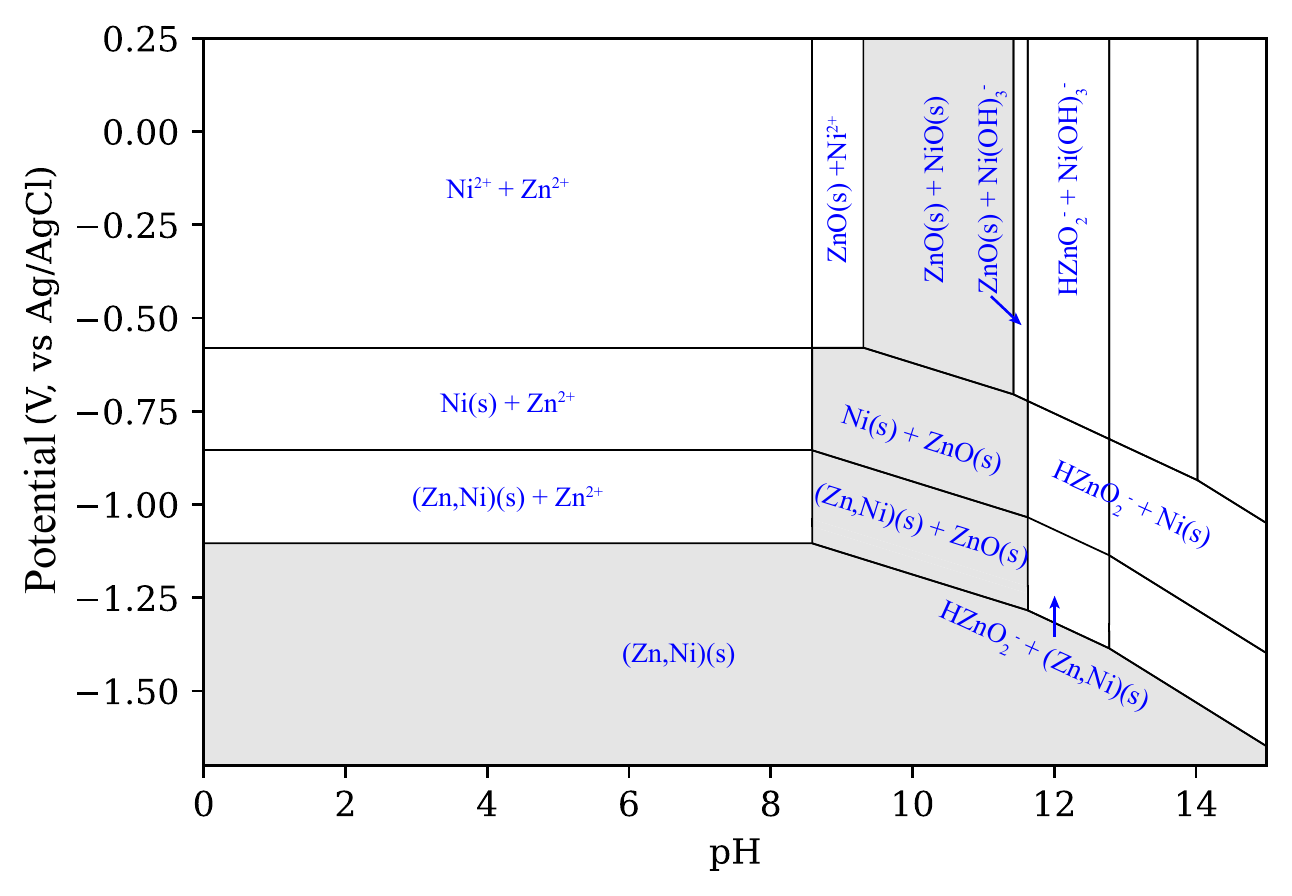}
    \caption{Pourbaix diagram generated from Materials Project data.  Specific regions attributed to stable intermetallic phases removed for simplicity.  The pourbaix diagram predicts a region of stability between $\approx$ 8.5 and 11.5 in pH.}
    \label{fig:pour}
\end{figure*}

Based on the XPS data we can begin to understand the reactions that are occurring during the OCP hold.  Zn alloys in general \cite{porter1994corrosion}, and Zn-Ni alloys specifically\cite{eliaz2010electroplating}, are known to form a native carbonate scale that is Zn rich.  As this scale grows the Ni is segregated from the near surface region, reducing its detection by XPS from its bulk composition of 15 \atp.   We can further gain insight into corrosion in this system by considering the Pourbaix diagram of this alloy generated from The Materials Project\cite{Jain2013,persson2012prediction} (assuming the composition and an ion concentration in the solution of $10^{-6}$ ions of each metal species): Fig \ref{fig:pour} . 
Given the proclivity of Zn to form carbonates, we also consider the Pourbaix diagram in the presence of carbonic acid by \citet{delahay1951potential}.  While both diagrams should be taken as approximations due to violation of underlying assumptions (e.g., for the Materials Project diagram: 0 K density functional theory basis for phases, missing phases including hydroxides, deaerated electrolyte), these diagrams can provide further information, including possible phase regions and their stability, to understand the reactions at the surface.  The materials project diagram suggests that the Zn will dissolve or react at all pHs at a lower potential than the Ni.  In both cases we also see that at low pH the metals will dissolve and at high pH they will form oxides or hydroxides that will similarly dissolve.  In moderate pHs near neutral the metals will form oxides or hydroxides that are stable, providing a reduction in the corrosion rate.  It is interesting to note that the Materials Project Pourbaix diagram for \ce{Zn85Ni15} has approximately the same passive region as as pure Zn due to the thermodynamic stability region of ZnO.

Taking this information together we can surmise that in low pH solution, the Zinc carbonate scale, followed by the underlying Zn metal, is rapidly dissolved, leaving behind a Ni enriched surface with a high OCP.   As the pH increases the driving force for dissolution is reduced so the surface remains more Zn rich, lowering its OCP.  The XPS results also suggest that the C is preferentially dissolving from the surface.  At a pH around 3.5 we begin to see the corrosion rate decrease and the Ni content near the surface fall below that of the untreated surface.  This implies that we are, in fact, growing a passivating surface oxide or carbonate at a lower pH than thermodynamically predicted, an observation common to a great many metallic systems.  This surface layer increases the OCP of the surface, particularly as pH increases.  Finally at around pH 11, the Zn oxide/carbonate layer becomes soluble, and the surface shifts towards the underlying metal, which has a much lower OCP.  We assume that in the time interval between when the surface was chemically treated and when the sample was placed in the UHV (ulta-high vacuum) of the chamber (around 18 h), the Zn at the surface was again able to react with the atmosphere to form a fresh native scale and enrich the surface.  Taken together, the corrosion data from the SDC is consistent with what is known about this system from literature and our XPS data.

There are several aspects of this system we hope to improve in the future.  The biggest hurdle is the high resistance between the working electrode and the reference electrode.  This drop makes measurements of high current processes such as cyclic voltammograms of the Zn-Ni plate difficult.  There are two solutions to this issue.  The first would be an alternate design of the cell head that allows for placing the reference electrode (and perhaps the counter electrode) closer to the sample surface and expanding the flow channels between the electrodes.  In reality the geometry of the cell head makes this difficult but we are investigating smaller reference electrodes for this purpose.  Alternatively a patterned coating could be used to reduce the area of the film being sampled by the electrochemical cell.  This would reduce the reaction area and consequently the current that needs to flow through the cell head.  There is also some issue with uniformity of the corrosion due to nonuniformities in the flow rate of the electrolyte over the surface.  Again this could be improved through further optimization of the cell head.

\section{Conclusion}
We have developed a highly automated system for rapid corrosion measurements on planar samples.  Here we have demonstrated its use on a Zn-Ni alloy at a range of pHs.  We show through two measurement types that this alloy has a high corrosion rate at both low ($<3.5$) and high ($>12$) pH, but due to formation of a passivating layer the corrosion rate is much lower near neutral pH.  Beyond the application of the SDC demonstrated here, we think the cell can be used to probe a variety of sample types including combinatorial thin-films with compositional or structural variation \citep{joress2020high} and additively manufactured samples with variation in composition\cite{moorehead2020high,dippo2021highthroughput} or processing parameters\citep{weaver2021demonstration}.  Further, the automated nature of the system (including the data-processing capabilities) can allow for more efficient mapping of electrolyte and/or sample space through addition of a machine learning agent\cite{rohr2020benchmarking}.  Beyond use of this system as a corrosion mapping tool, we are developing it to also be used to deposit alloys through electroplating. In this way we can create a fully closed loop platform capable of on-demand synthesis and corrosion to accelerate materials discovery\cite{aspuru2018materials}.

\section*{Supplementary Information}
A PDF document with figures, as noted in the text, is attached.  In addition the entire data-set along with code for data handling and plotting can be found here: https://github.com/usnistgov/ZnNi-pH-series
Underlying analysis and SDC control code can be found at https://github.com/USNISTGOV/auto-sdc and https://github.com/usnistgov/tafel-fitter

\section*{Acknowledgements}
The Zn-Ni sample was provided by Atotech Inc. The authors acknowledge D. Josell (NIST) for useful discussion and C. Amigo (NIST) for machining the cell head and other components. NH acknowledges funding from Advanced Research Projects Agency-Energy (ARPA-E) award number DEAR0001050.

\bibliographystyle{elsarticle-num-names} 
\bibliography{refs}





\end{document}